\documentclass[10pt,a4paper]{article}
\usepackage[utf8]{inputenc}
\usepackage{amsmath}
\usepackage{amsfonts}
\usepackage{amssymb}
\usepackage{graphicx}
\usepackage{graphics}
\usepackage{epstopdf}
\usepackage{cite}
\usepackage{subfig}
\usepackage{float}
\graphicspath{{img/}}
\usepackage[toc,page]{appendix}
\usepackage{geometry}
\usepackage{authblk}
\usepackage{abstract}
\title{Oscillatory amplitude of stochastic gravitational wave spectrum}
\author[1]{N. Malsawmtluangi \thanks{ Corresponding author, E-mail: \texttt{tei.naulak@uohyd.ac.in}}}
      
\author[2]{P. K. Suresh}
\affil[1,2]{\small School of Physics, University of Hyderabad\\ P.O. Central University, Hyderabad 500046. India.}
\affil[1]{\small Department of Physics, Government Kolasib College, Kolasib 796081, Mizoram, India.}
\date{}

\begin{document}
\maketitle

\begin{abstract}
Primordial gravitational waves generated from early universe are placed in the squeezed vacuum state and the resulting stochastic background is studied for various models of the expanding universe. The quantum effect   on the   stochastic gravitational waves leads to overall enhancement  of the amplitude and spectral energy density when compared to those in the absence of squeezing effect with continued increase in the amplitude in  the accelerating stage  and oscillatory  behavior  at  higher frequency range of the spectrum  in the accelerating universe. Through the quantum effect, it is also found that the reheating phenomenon affects the entire spectrum. The results of the present study  may be useful to  test the possibility of detection of the stochastic  gravitational waves by current and future gravitational wave detectors and whether these waves exist in the squeezed vacuum state.
\end{abstract}

\noindent
{\bf keywords}: gravitational waves; accelerating universe, squeezed state, stochastic background, early universe


\newpage

\section{Introduction}
The existence of gravitational waves is one of the finest predictions of the theory of general relativity. Gravitational waves can be generated by various sources, from quantum fluctuations during inflationary expansion in the very early universe to astrophysical sources like black hole mergers, binary systems of neutron stars and pulsars, core collapse of supernova, to name a few \cite{bss, rhb, adl}. The direct detections of gravitational waves to date have all been from black hole mergers, neutron star mergers or black hole-neutron star binaries \cite{gw1, gw2, gw3, gw4, gw5, gw6, gw7, gw8} while the indirect detections have been from binary systems of neutron stars and pulsars \cite{ht, mb}  and  the gravitational waves  from the early universe, called primordial or relic gravitational waves, are yet to be detected \cite{pgw1, pgw2, pgw3, pgw4}.

Gravitational waves, in general, are very important in understanding the dynamics of the universe and various astrophysical phenomena. Since they are very weak, they couple very weakly to matter and travel almost unimpeded since their generation. As such, they carry vital information about their source almost unimpeded, hence they provide the new window to explore the universe, and bring to light those parts which cannot be probed via other methods like light and other electromagnetic waves. They can also provide the definitive test for general theory of relativity and its modifications or extended theories of gravity like $f(R)$ theories \cite{fr1, fr2, fr3}. Thus, gravitational waves can shed light to processes in the universe which are otherwise difficult or impossible to observe with other means.

Primordial gravitational waves are believed to have been generated during the very early stage of the universe from quantum fluctuations during the inflationary process \cite{lp1, sns}. They are believed to have traversed the universe through its various evolutionary stages, from the inflationary period upto the current accelerating stage with an ever increasing wavelength with the expansion of the universe. The  primordial gravitational waves are expected to form a stochastic background of standing waves whose spectrum depends on and varies with each evolutionary stage of the universe. These primordial gravitational waves are believed to carry important information about the very early universe and its very own formation.

Due to the quantum nature of the generation process of the primordial gravitational waves, the primordial gravitational waves themselves are expected to also be in a specific quantum state called the squeezed vacuum state \cite{lp2}. This resulted from the evolution of the initial vacuum state into a multi-particle quantum state which is induced by the strong and variable gravitational field \cite{lp3, lp4, lp5, lp6}. Due to  the squeezing effect, the variance of the wave mode's phase is  squeezed while its amplitude is increased, thus  the corresponding uncertainty product remains intact. The associated squeezing parameter and the mean number of quanta  of the gravitational waves are growing functions in the amplifying regime  of the early universe and  stop growing only at the end of  it. Thus, if the primordial gravitational waves are indeed in the squeezed vacuum state, then this squeezing effect is expected to be reflected on the spectral amplitude  of the stochastic background of the primordial gravitational waves.  Therefore, from this observational point of view, it is important to  study  the  amplitude of the stochastic background of primordial gravitational waves  in the squeezed vacuum state with  the  sensitivity of various gravitational wave detectors, hence the present work.

In the present work, we obtain the spectrum of primordial gravitational waves  in the squeezed vacuum state for the  flat Friedmann-Lemaitre-Robertson-Walker (FLRW) universe including  the  current accelerating  phase which was not done in earlier works on squeezed primordial gravitational waves. The  amplitude of the spectrum is studied   with  the sensitivity curves of  various  ongoing and proposed detectors  \cite{d1, d2, d3, d4, d5, d6, d7, d8, d9, d10, d11, d12}. The amplitude of the  stochastic gravitational waves in the squeezed vacuum state  is found to be enhanced by several amount compared to the non-squeezed case and oscillatory  at higher frequency which  are  the reflections of  the quantum effect.

\section{Expansion history of the universe}
According to the standard model of cosmology, the universe, throughout its expansion, has undergone several evolutionary stages. The scale factor $a$ characterizes the relative expansion of the universe with time. The different evolutionary stages of the universe can, therefore, be represented in terms of the scale factor for each corresponding range of conformal time $\varsigma$ as follows \cite{lp2, cqg}.

For the initial inflationary  stage, also called the $i$-stage,
\begin{equation}\label{inf}
a(\varsigma) = l_0|\varsigma|^{1+\beta},  ~~~~ -\infty < \varsigma \leqslant \varsigma_1,
\end{equation}
where $l_0$ and $\beta$ are arbitrary constants, $\beta < -1$ and $\varsigma_1 < 0$. $d\varsigma=dt/a$ is the conformal time and $t$ is the proper time.

The inflationary index $\beta$ is related to the equation of state during inflation. The inflationary stage is governed by matter with the effective equation of state $p= w \epsilon$, $p$ being the effective pressure of matter and $\epsilon$ being the energy density, where $w = (1-\beta)/3(1+\beta)$ and $w$ varies from $-1/3$ to $-\infty$ for $-\infty < \beta < -1$ \cite{lp7, jm}. All expanding models with $\beta < -1$ are inflationary. The cases $\beta < -2$ correspond to the power-law inflation $a(t) \sim t^m$, $m>1$, $t>0$ while the case with $\beta = -2$ corresponds to the de Sitter expansion with $w=-1$ and the cases $-2<\beta <-1$ correspond to the law of expansion $a(t) \sim |t|^m$, $m<-1$, $t<0$. For the case $-2<\beta <-1$, expansion is accompanied by the growth of energy density and curvature. Further constraint on $\beta$ is given in later sections.
 
For the reheating stage, denoted by the subscript $z$,
\begin{equation}\label{azd}
a(\varsigma)= a_z (\varsigma- \varsigma_p)^{1+\beta_s}, ~~~~ \varsigma_1 \leqslant \varsigma \leqslant \varsigma_s,
\end{equation}
where $\beta_s$ describes the expansion behavior of the reheating stage.

The radiation-dominated stage, denoted by the subscript $e$ has the scale factor,
\begin{equation}\label{ard}
a(\varsigma)= a_e(\varsigma-\varsigma_e),  ~~~~ \varsigma_s \leqslant \varsigma \leqslant \varsigma_2.
\end{equation}

For the matter-dominated stage, denoted by the subscript $m$,
\begin{equation}\label{amd}
a(\varsigma)= a_m (\varsigma - \varsigma_m)^2,  ~~~~ \varsigma_2 \leqslant \varsigma \leqslant \varsigma_E.
\end{equation}

The accelerating stage upto the current stage, denoted by the subscript $a$ has,
\begin{equation}\label{acc}
a(\varsigma)=l_H |\varsigma -\varsigma_a |^{-1},  ~~~~ \varsigma_E \leqslant \varsigma \leqslant \varsigma_H,
\end{equation}
where $l_H$ is the present-day Hubble radius,
\[l_H=\left(\frac{a^2}{a'}\right) _{\varsigma_H} = \frac{1}{H},\]
where prime ($'$) indicates the derivative with respect to the conformal time $\varsigma$. Eqs.(\ref{inf})-(\ref{acc}) give the evolution of the universe in terms of the scale factor where $\varsigma_1$, $\varsigma_s$, $\varsigma_2$ and $\varsigma_E$ are the conformal time denoting the various evolutionary stages of the universe and mark the points of successive transition from one stage to the next. By taking the normalization $|\varsigma_H - \varsigma_a |=1$, we get the link between these instances of conformal time with the help of the continuous joining of the scale factor and its derivatives at each point:
\begin{eqnarray}
\varsigma_E - \varsigma_a &=& \xi_E, \nonumber \\
\varsigma_E - \varsigma_m &=& 2\xi_E, \nonumber \\
\varsigma_2 - \varsigma_m &=& 2\xi_E \xi_2^{-1/2}, \nonumber \\
\varsigma_2 - \varsigma_e &=& \xi_E \xi_2^{-1/2}, \nonumber \\
\varsigma_s - \varsigma_e &=& \xi_E \xi_2^{-1/2}\xi_s^{-1}, \nonumber \\
\varsigma_s - \varsigma_p &=& | 1+\beta_s | \xi_E \xi_2^{-1/2}\xi_s^{-1}, \nonumber \\
\varsigma_1 - \varsigma_p &=& | 1+\beta_s |\xi_E \xi_2^{-1/2}\xi_s^{-1} \xi_1^{-(1/1+\beta_s)}, \nonumber \\
\varsigma_1  &=& | 1+\beta |\xi_E \xi_2^{-1/2}\xi_s^{-1} \xi_1^{-(1/1+\beta_s)},
\end{eqnarray}
and the constants at each stage:
\begin{eqnarray}
a_m &=& \frac{l_H}{4}\xi_E^{-3}, \nonumber \\
a_e &=& l_H \xi_E^{-2}\xi_2^{-1/2}, \nonumber \\
a_z &=& l_H |1+\beta_s|^{-(1+\beta_s)}\xi_E^{-(2+\beta_s)}\xi_2^{(\beta_s-1)/2}\xi_s^{\beta_s}, \nonumber \\
l_0 &=& l_H b\xi_E^{-(2+\beta)}\xi_2^{(\beta-1)/2}\xi_s^{\beta}\xi_1^{(\beta-\beta_s)/(1+\beta_s)}, \label{lo}
\end{eqnarray}
where $b=|1+\beta|^{-(1+\beta)}$, and $\xi_E\equiv a(\varsigma_H)/a(\varsigma_E)$, $\xi_2\equiv a(\varsigma_E)/a(\varsigma_2)$, $\xi_s\equiv a(\varsigma_2)/a(\varsigma_s)$, $\xi_1\equiv a(\varsigma_s)/a(\varsigma_1)$.

In the accelerating universe  driven by dark energy, taking the dark energy and dark matter densities as $\Omega_{\Lambda} \sim 0.7$ and $\Omega_m \sim 0.3$ respectively, the redshift corresponding to $\varsigma_E$, the time of equality of dark energy density and dark matter density, is given by,
\begin{eqnarray}
\frac{a(\varsigma_H)}{a(\varsigma_E)} = (1 + z_E) = \left(\frac{\Omega_{\Lambda}}{\Omega_m}\right)^{1/3} \simeq 1.33.
\end{eqnarray}
The matter dominated stage started at the time instant $\varsigma_2$ for which, according to the Planck observational data, the corresponding redshift is $\sim 3371$ \cite{pcb}. Therefore,
\begin{eqnarray}
\frac{a(\varsigma_E)}{a(\varsigma_2)} = (1 + z_2) \simeq 3371.
\end{eqnarray}
The radiation dominated stage spanned from $\varsigma_s$ to $\varsigma_2$. The corresponding temperatures at the starting and ending moments are assumed to be respectively $T_s \simeq 10^{15}$ GeV and $T_2 \simeq 1$ eV, where the starting temperature is assumed to be the typical energy scale of the GUT era \cite{cqg2}. Therefore,
\begin{eqnarray}
\frac{a(\varsigma_2)}{a(\varsigma_s)} = \frac{T_s}{T_2} = 10^{24}.
\end{eqnarray}
The reheating stage which spanned from $\varsigma_1$ to $\varsigma_s$ is simply a general expansion stage as this stage has not been properly understood. It could have been governed by a matter either stiffer or softer than radiation. It could also be part of radiation dominated stage, as for a particular choice $\beta_s = 0$, this stage reduces to that of radiation dominated. Thus, for definiteness in computation, a simple choice is made for $\xi_1 = \frac{a(\varsigma_s)}{a(\varsigma_1)} = 300$ for which the implications of the choice are not in conflict with observations.

\section{Primordial gravitational waves}
The perturbed metric of a flat FLRW universe can be written as,
\begin{equation}
dS^2 = a^2(\varsigma)[-d\varsigma^2 + (\delta_{ij}+h_{ij})dx^idx^j],
\end{equation}
where $h_{ij}$ is a transverse-traceless perturbation of space-time such that $|h_{ij}|\ll\delta_{ij}$,
~~~\[\partial_ih^{ij} = 0 , ~~~\delta^{ij}h_{ij}=0,\]
where $\delta_{ij}$ is the flat space metric.

The gravitational wave field $h_{ij}(\textbf{x},\varsigma)$ can be expanded over spatial Fourier harmonics $e^{\pm i\textbf{k}.\textbf{x}}$, where $\textbf{k}$ is a wave vector. Thus,
\begin{equation}\label{fourm}
h_{ij}(\textbf{x},\varsigma) = \frac{D}{(2\pi)^{3/2}}\int_{-\infty}^{+\infty}\frac{d^3\textbf{k}}{\sqrt{2k}}\sum_{p=1}^2 [h_k^{(p)}(\varsigma) c_k^{(p)} \varepsilon_{ij}^{(p)}(\textbf{k})e^{i\textbf{k}.\textbf{x}} + h_k^{(p) \ast}(\varsigma)c_k^{(p) \dagger} \varepsilon_{ij}^{(p) \ast}(\textbf{k})e^{-i\textbf{k}.\textbf{x}}],
\end{equation}
where $D=\sqrt{16\pi}l_{pl}$ is the normalization constant, $l_{pl}=\sqrt{G}$ is Planck's length. This particular value of $D$ guarantees correct quantum normalization of the field: energy $\hbar \omega/2$ per each mode in the initial vacuum state. Of course, we have taken $\hbar=1$ throughout this paper. The wave number is $k=(\delta_{ij}k^ik^j)^{1/2}$ and is related to wavelength $\lambda$ by $\lambda = 2\pi a/k$.

The two linear polarization states $\varepsilon_{ij}^{(p)}$, where $p=1,2$, are symmetric and transverse-traceless and satisfy the conditions
\[\varepsilon_{ij}^{(p)}\delta^{ij}=0, ~~\varepsilon_{ij}^{(p)}k^i=0, ~~\varepsilon_{ij}^{(p)}\varepsilon^{ ij(p')} = 2\delta_{pp'}, ~~\varepsilon_{ij}^{(p)}(\textbf{-k})=\varepsilon_{ij}^{(p)}(\textbf{k}).\]
These polarizations states are called the plus $(+)$ polarization and cross $(\times)$ polarization. The contributions from both these polarizations are same, therefore the index $p$ is dropped from here onward for convenience.

The creation and annihilation operators $c_k^{\dagger}$ and $c_k$ satisfy the relationships
\[[c_k,c_{k'}^{ \dagger}]=\delta^3(k-k'),\]
\[[c_k,c_{k'}]=[c_k^{\dagger},c_{k'}^{ \dagger}]=0.\]
The Heisenberg equations of motion govern the evolution of annihilation and creation operators as,
\begin{equation}\label{Hsa}
\frac{d}{d\varsigma}c_{{k}}(\varsigma) = -i[c_k(\varsigma),H_{gw}],
\end{equation}
\begin{equation}\label{Hsb}
\frac{d}{d\varsigma}c_{{k}}^{\dagger}(\varsigma) = -i[c_k^{\dagger}(\varsigma),H_{gw}],
\end{equation}
where $H_{gw}$ is the Hamiltonian for primordial gravitational waves. See appendix A for more details.

The initial vacuum state $|0\rangle$ is defined as
\[c_k |0\rangle = 0.\]
The Bogoliubov transformations for Eq.\eqref{Hsa} and Eq.\eqref{Hsb} are
\begin{equation}
c_{{k}}(\varsigma) = u_{k}(\varsigma)c_{{k}}(0) + v_{k}(\varsigma)c_{{k}}^{\dagger}(0),
\end{equation}
\begin{equation}
c_{{k}}^{\dagger}(\varsigma) = u_k^{\ast}(\varsigma)c_{{k}}^{\dagger}(0) + v_k^{\ast}(\varsigma)c_{{k}}(0),
\end{equation}
where $c_k(0)$ and $c_k^{\dagger}(0)$ are the initial values of the operators and $u_k(\varsigma)$ and $v_k(\varsigma)$ are complex functions. These functions satisfy the condition
\[|u_k|^2-|v_k|^2=1.\] 
The dynamical evolution equation of the primordial gravitational waves in the flat FLRW universe can be written as,
\begin{equation}
h''_k (\varsigma) + 2\frac{a'}{a} h'_k (\varsigma) + k^2h_k (\varsigma) = 0.
\end{equation}
The gravitational wave mode $h_k (\varsigma)$ can be rescaled in terms of mode function as,
\begin{equation}\label{mf}
h_k(\varsigma) a(\varsigma)= \mu_k(\varsigma),
\end{equation}
where the mode functions can have the following form,
\begin{equation}\label{mff}
\mu_k(\varsigma) = u_k(\varsigma) + v_k^{\ast}(\varsigma),
\end{equation}
which then satisfies the equation of  motion
\begin{equation}\label{eom}
\mu_k'' (\varsigma) + \left(k^2-\frac{a''}{a}\right)\mu_k (\varsigma) = 0.
\end{equation}
There can be two limiting cases for Eq.\eqref{eom}: $k^2\gg a''/a$ and $k^2\ll a''/a$.
The limit $k^2\gg a''/a$ indicates the short wavelength limit where the wave is outside the potential barrier and does not interact with the barrier and propagates with an adiabatically decreasing amplitude $h_k(\varsigma) \propto 1/a(\varsigma)$.
The limit $k^2\ll a''/a$ indicates the long wavelength limit where the wave is inside the potential barrier. The wave interacts with the barrier and gets parametrically amplified above $h_k(\varsigma) \propto 1/a(\varsigma)$ and due to this, the initial vacuum state of these modes is transformed into a multi-particle quantum state called the squeezed vacuum state.

\subsection{Primordial gravitational waves in the squeezed vacuum state}

The complex functions $u_k(\varsigma)$ and $v_k (\varsigma)$ in Eq.\eqref{mff} can be represented in terms of  the squeezing parameter $r_k$, squeezing angle $\phi_k$ and the rotation angle $\theta_k$ as \cite{lp2},
\begin{eqnarray} \label{y}
u_k &=& e^{i \theta_k} \cosh r_k, \nonumber \\
v_k &=& e^{-i(\theta_k -2\phi_k)} \sinh r_k.
\end{eqnarray}
The equations of motion for these two complex functions are,
\begin{eqnarray}
i\frac{du_k}{d\varsigma} &=& ku_k + i\frac{a'}{a} v^{\ast}_k, \nonumber \\
 i\frac{dv_k}{d\varsigma} &=& kv_k + i\frac{a'}{a} u^{\ast}_k,
\end{eqnarray}
which lead to the equations governing the three real functions mentioned above:
\begin{eqnarray}\label{aaa}
r'_k &=&\frac{a'}{a}\cos 2\phi_k, \nonumber \\  
\phi'_k &=& -k-\frac{a'}{a}\sin 2\phi_k \coth 2r_k,\\ 
\theta'_k &=& -k-\frac{a'}{a} \sin 2\phi_k \tanh r_k. \nonumber
\end{eqnarray}
The two-point correlation function of the gravitational wave field defines the power spectrum of the gravitational waves as,
\begin{equation}\label{2pt}
\langle0|h_{ij}(\textbf{x},\varsigma)h^{ij}(\textbf{x},\varsigma)|0\rangle = \frac{D^2}{2\pi^2}\int_{0}^{\infty}k^2 |h_k(\varsigma)|^2\frac{dk}{k},
\end{equation}
where
\begin{equation}\label{amp2}
h^2(k,\varsigma)=\frac{D^2}{2\pi^2}k^2|h_k(\varsigma)|^2
\end{equation}
gives the mean-square value of the gravitational waves with interval $k$.
Also,
\begin{equation}
h^2(k,\varsigma) =\frac{1}{2}|h(k,\varsigma)|^2,
\end{equation}
where,
\begin{equation}\label{e}
|h(k,\varsigma)|=\frac{D}{\pi}k|h_k(\varsigma)|.
\end{equation}
Using $D = \sqrt{16\pi}l_{pl}$, we get the power spectrum as 
\begin{equation}\label{psp}
|h(k,\varsigma)|=\frac{4l_{pl}}{\sqrt{\pi}}k|h_k(\varsigma)|.
\end{equation}
Using Eqs.\eqref{mff} and \eqref{y},
\begin{equation}\label{sqz}
|h_k(\varsigma)|^2 = \frac{1}{a^2(\varsigma)}\left(1 + 2\sinh^2r_k + \sinh 2r_k \cos 2 \phi_k \right).
\end{equation}
Considering the initial condition as the $i$-stage, the wavelength of a wave with wave number $k$ which crossed over the horizon at time $\varsigma_i$ is
\begin{equation}\label{wvl}
\lambda_i=\frac{2\pi a(\varsigma_i)}{k} = \frac{1}{H(\varsigma_i)},
\end{equation}
and Eq.\eqref{inf} gives
\begin{equation}
\frac{1}{H(\varsigma_i)}=\frac{l_0|\varsigma_i|^{2+\beta}}{|1+\beta|}.
\end{equation}
Using Eqs.\eqref{psp}, \eqref{sqz} and \eqref{wvl}, we get
\begin{equation}\label{qp}
h(k,\varsigma) = 8\sqrt{\pi}\frac{l_{pl}}{\lambda_i}(1 + 2\sinh^2r_k + \sinh 2r_k \cos 2 \phi_k)^{1/2}.
\end{equation}
Suppose the initial condition of the mode function, i.e., before the squeezing effect takes place, is
\begin{equation}
|h_k(\varsigma_i)|=\frac{1}{a(\varsigma_i)},
\end{equation}
obtained from the solution of the mode function equation in Eq.\eqref{eom}. Then from Eqs.\eqref{psp} and \eqref{wvl}, we get
\begin{equation}\label{qn}
h(k,\varsigma) = 8\sqrt{\pi}\frac{l_{pl}}{\lambda_i}.
\end{equation}
The wave number corresponding to the current Hubble radius is,
\begin{equation}
k_H = \frac{2\pi a(\varsigma_H)}{l_H},
\end{equation}
which leads to
\begin{equation}\label{l}
\lambda_H =\frac{k_H l_H}{k}.
\end{equation}
Thus, the amplitude of the primordial gravitational waves for the frequency ranges corresponding to the inflation upto the current accelerating stage can be expressed as,
\begin{equation}\label{hkbe}
h(k,\varsigma_H)= 8\sqrt{\pi}\left(\frac{l_{pl}}{l_H}\right)\left(\frac{k}{k_H}\right)(1 + 2\sinh^2r_k + \sinh 2r_k \cos 2 \phi_k)^{1/2},
\end{equation}
where $k_H$ is the wave number corresponding to the present time. The term $(1 + 2\sinh^2r_k + \sinh 2r_k \cos 2 \phi_k)$ represents the squeezing effect.

\subsection{Squeezing parameter and squeezing angle}
The squeezing parameter and squeezing angle are time dependent, they grow with time, therefore their values vary for each frequency range in the expanding universe. 

In the adiabatic regime, the wavelength is shorter than the Hubble radius, therefore $k$ is dominant. Thus the squeezing angle can be given by,
\begin{equation}
\phi_k = -k(\varsigma + \varsigma_k) = -\frac{k}{a(\varsigma)}\left(1+\frac{a_{\ast}(\varsigma)}{a(\varsigma_k)}\right),
\end{equation}
where $\varsigma_k$ is constant. The evaluation is done at present time where $a(\varsigma) \propto \varsigma^{-1}$. In order to find $a_{\ast}(\varsigma)/a(\varsigma_k)$, we consider $a(\varsigma_k)$ at $a(\varsigma_H)$ and $a_{\ast}(\varsigma)$ as the conformal time at the starting point of time range. 
From Eqs.\eqref{azd}-\eqref{acc}, one can write the rough relationship between the scale factor and conformal time for each evolutionary stage as,
\begin{eqnarray}\label{vs}
&& \varsigma_1 \leqslant \varsigma \leqslant \varsigma_s \rightarrow a(\varsigma) \propto \varsigma^{1+\beta_s}, \nonumber \\
&& \varsigma_s \leqslant \varsigma \leqslant \varsigma_2 \rightarrow a(\varsigma) \propto \varsigma, \nonumber \\
&& \varsigma_2 \leqslant \varsigma \leqslant \varsigma_E \rightarrow a(\varsigma) \propto \varsigma^2, \nonumber \\
&& \varsigma_E \leqslant \varsigma \leqslant \varsigma_H \rightarrow a(\varsigma) \propto \varsigma^{-1}. 
\end{eqnarray}
Now, consider $a_{\ast \ast}(\varsigma)$, the conformal time at the end of time range. For every given wave number $k$, the quantity $a_{\ast \ast}$ is determined by the condition $\lambda (\varsigma_{\ast \ast}) = l (\varsigma_{\ast \ast})$, whereas  the quantity $a_\ast$ is determined by the condition $\lambda (\varsigma_\ast) = l (\varsigma_\ast)$, where $l (\varsigma_{\ast \ast})$ and $l (\varsigma_\ast)$ are the Hubble radius $l(\varsigma) = a^2/a'$ at $\varsigma_{\ast \ast}$ and $\varsigma_\ast$ respectively and $\lambda (\varsigma_{\ast \ast})$ and $\lambda (\varsigma_\ast)$ are the corresponding wavelengths, $\lambda = 2\pi a(\varsigma)/k$. Considering the range $k_s \leqslant k \leqslant k_1$ as an example, from Eq.\eqref{azd}, it can be checked that
\begin{equation}
\xi_1 = \frac{a(\varsigma_s)}{a(\varsigma_1)} = \frac{(\varsigma_s-\varsigma_p)^{1+\beta_s}}{(\varsigma_1-\varsigma_p)^{1+\beta_s}}.
\end{equation}
The Hubble radius corresponding to the conformal time $\varsigma_s$ and $\varsigma_1$ can be found as,
\begin{eqnarray}
l(\varsigma_s) &=& (\varsigma_s-\varsigma_p)^{2+\beta_s}, \nonumber \\
l(\varsigma_1) &=& (\varsigma_1-\varsigma_p)^{2+\beta_s}.
\end{eqnarray}
Taking their ratio,
\begin{eqnarray}\label{rl}
\frac{l(\varsigma_s)}{l(\varsigma_1)} &=& \frac{(\varsigma_s-\varsigma_p)}{(\varsigma_1-\varsigma_p)}\frac{(\varsigma_s-\varsigma_p)^{1+\beta_s}}{(\varsigma_1-\varsigma_p)^{1+\beta_s}}, \nonumber \\
 &=& \frac{a(\varsigma_s)}{a(\varsigma_1)} \left[\frac{a(\varsigma_s)}{a(\varsigma_1)}\right]^{1/1+\beta_s}.
\end{eqnarray}
The ratio of the corresponding wavelengths is
\begin{equation}\label{rlb}
\frac{\lambda(\varsigma_s)}{\lambda(\varsigma_1)} = \frac{k_1}{k_s}\frac{a(\varsigma_s)}{a(\varsigma_1)}.
\end{equation}
Then, from Eqs.\eqref{rl} and \eqref{rlb}, we get,
\begin{equation}\label{vs1}
\frac{k_1}{k_s} = \left[\frac{a(\varsigma_s)}{a(\varsigma_1)}\right]^{1/1+\beta_s} = \xi_1^{1/1+\beta_s}.
\end{equation}
Following in the same method, we get,
\begin{eqnarray}\label{vs2}
&& \xi_1 = \left(\frac{k_1}{k_s}\right)^{1+\beta_s},~~~~~~~\xi_s = \frac{k_s}{k_2}, \nonumber \\
&& \xi_2 = \left(\frac{k_2}{k_E}\right)^2, ~~~~~~~~~~\xi_E = \left(\frac{k_E}{k_H}\right)^{-1}.
\end{eqnarray}
Then, with the help of Eq.\eqref{vs}, we have,
\begin{eqnarray}\label{vs4}
\frac{a_{\ast}(\varsigma)}{a(\varsigma_k)} &=& \frac{a_{\ast}(\varsigma)}{a(\varsigma_s)} \frac{a(\varsigma_s)}{a(\varsigma_2)}  \frac{a(\varsigma_2)}{a(\varsigma_E)}  \frac{a(\varsigma_E)}{a(\varsigma_H)}, \nonumber \\
&=& \left(\frac{k_E}{k}\right)\left(\frac{k_E}{k_2}\right)\left(\frac{k_s}{k}\right)^{\beta_s}\left(\frac{k_E}{k_H}\right),
\end{eqnarray}
where, for successive intervals,
\begin{eqnarray}\label{vs3}
&& a_{\ast}(\varsigma) \propto \varsigma^{1+\beta_s} \propto k^{-(1+\beta_s)},~~~~~a(\varsigma_s) \propto \varsigma_s^{1+\beta_s} \propto k_s^{-(1+\beta_s)} \nonumber \\
&& a(\varsigma_s) \propto \varsigma_s \propto k_s^{-1}, ~~~~~~~~~~~~~~~~~a(\varsigma_2) \propto \varsigma_2 \propto k_2^{-1}, \nonumber \\
&& a(\varsigma_2) \propto \varsigma_2^2 \propto k_2^{-2}, ~~~~~~~~~~~~~~~~a(\varsigma_E) \propto \varsigma_E^2 \propto k_E^{-2}, \nonumber \\
&& a(\varsigma_E) \propto \varsigma_E^{-1} \propto k_E, ~~~~~~~~~~~~~~~a(\varsigma_H) \propto \varsigma_H^{-1} \propto k_H.
\end{eqnarray}
Strictly speaking, $a(\varsigma)$ and $k$ are not inversely proportional to each other, but their ratios are, following Eq.\eqref{vs1} and similar evaluations which lead to Eqs.\eqref{vs2}. The relations in Eqs.\eqref{vs3} are simply a breakdown of the numerators and denominators in Eq.\eqref{vs4}.

In the long wavelength regime, $k$ can be neglected from Eq.\eqref{aaa}, and the squeezing angle becomes,
\begin{equation}
\phi_k \propto \tan^{-1} \left(\frac{1}{a^2(\varsigma)}\right).
\end{equation}
As such, we calculated the squeezing angle for each frequency range as:
\begin{eqnarray}
\phi_k &=& -k\left[1+\left(\frac{k_E}{k}\right)\left(\frac{k_E}{k_2}\right)\left(\frac{k_s}{k}\right)^{\beta_s}\left(\frac{k_E}{k_H}\right)\right], ~~~ k_s \leqslant k \leqslant k_1, \label{sqa1} \\
\phi_k &=& -k\left[1+\left(\frac{k_E}{k}\right)\left(\frac{k_E}{k_2}\right)\left(\frac{k_E}{k_H}\right)\right], ~~k_2 \leqslant k \leqslant k_s, \label{sqa2}\\
\phi_k &=& -k\left[1+\left(\frac{k_E}{k}\right)^2\left(\frac{k_E}{k_H}\right)\right], ~~k_H \leqslant k \leqslant k_2, \label{sqa3}\\
\phi_k &=& \tan^{-1}(k_H^2),~~k_E \leqslant k \leqslant k_H, \label{sqa4}\\
\phi_k &=& \tan^{-1}(k_E^2), ~~ k \leqslant k_E. \label{sqa5}
\end{eqnarray}
For each of these squeezing angles $\cos \phi_k =1$ in each range such that $r_k'=a'/a$ in Eq.\eqref{aaa}. Hence, after integrating this, the frequency dependent squeezing parameter $r_k$ grows as,
\begin{equation}
r_k \approx \ln \frac{a_{\ast\ast}(k)}{a_{\ast}(k)},
\end{equation}
where $a_{\ast}$ is the value of $a(\varsigma)$ at $\varsigma_{\ast}$, the beginning of the time range, i.e., the higher end of the frequency range and $a_{\ast\ast}$ denotes $a(\varsigma)$ at $\varsigma_{\ast\ast}$, the end of time range, i.e., the lower end of frequency range.

All modes start in the vacuum state $|0\rangle$, i.e., $r_k =0$ initially. For the high frequency mode $k = k_1$, $a_{\ast} = a_{\ast\ast} = a(\varsigma_1)$ which gives $r_k = 0$. Thus the high frequency modes $k > k_1$ are not in the amplifying regime. The amplifying regime only starts at $k = k_1$. 

Since the evolution of the amplitude for each stage depends directly on the evolution of $r_k$ from Eq.\eqref{hkbe}, one needs to find $r_k$ for each stage. In order to find the present-day value of $r_k$, it is necessary to find the ratio $a_{\ast\ast}(k)/a_{\ast}(k)$ at each point of transition. This can be done with the help of Eqs.\eqref{vs2} and the steps leading to them.
The squeezing parameter $r_k$ can then be calculated in descending order of wave number by finding $a_{\ast \ast}(k)/a_\ast(k)$ and then multiplying it by the previous interval's $e^{r_k}$. Let us again take the range $k_s \leqslant k \leqslant k_1$ as an example as $e^{r_k}=1$ for $k_1 < k$. Then,
\begin{eqnarray}
\frac{a_\ast(k)}{a_\ast(k_1)} &=& \left(\frac{\varsigma}{\varsigma_1}\right)^{1+\beta} = \left(\frac{k_1}{k}\right)^{1+\beta}, \nonumber \\
\frac{a_{\ast \ast}(k)}{a_{\ast \ast}(k_s)} &=& \left(\frac{\varsigma}{\varsigma_s}\right)^{1+\beta_s} = \left(\frac{k_s}{k}\right)^{1+\beta_s}. \nonumber
\end{eqnarray}
Here, $a_{\ast \ast}(k_s) = a(\varsigma_s)$ and $a_\ast(k_1) = a(\varsigma_1)$. Then,
\begin{eqnarray}
\frac{a_{\ast \ast}(k)}{a_\ast(k)} = \left(\frac{k_1}{k_s}\right)^{1+\beta_s} \left(\frac{k_s}{k}\right)^{1+\beta_s} \left(\frac{k}{k_1}\right)^{1+\beta} \times \left[{\rm previous~interval's}~e^{r_k}\right]. \nonumber
\end{eqnarray}
Following in a similar manner, the squeezing parameter for each frequency range (in descending order) for the short-wavelength regime, $k\geqslant k_H$ becomes,
\begin{eqnarray}
r_k &=& \ln \left(\frac{k}{k_1}\right)^{\beta - \beta_s}, ~~~ k_s \leqslant k \leqslant k_1, \label{rk1} \\
r_k &=& \ln \left[\left(\frac{k}{k_s}\right)^{\beta}\left(\frac{k_s}{k_1}\right)^{\beta - \beta_s} \right], ~~k_2 \leqslant k \leqslant k_s, \label{rk2}\\
r_k &=& \ln \left[ \left(\frac{k}{k_2}\right)^{\beta - 1}\left(\frac{k_2}{k_1}\right)^{\beta} \left(\frac{k_s}{k_1}\right)^{-\beta_s} \right], ~~k_H \leqslant k \leqslant k_2. \label{rk3}
\end{eqnarray}
For the modes in the long wavelength regime, for $a_{\ast \ast} (k)$, one has to take $a(\varsigma_R)$, i.e., at reception, the frequency corresponding to the conformal time at the starting time of the range. Consider the range $k_E \leqslant k \leqslant k_H$ as an example. Then, $a_{\ast \ast} (k) = a_\ast (k_H)$ and divide both sides by $a_\ast (k)$. Here,
\begin{eqnarray}
\frac{a_\ast (k)}{a_\ast(k_H)} &=& \left(\frac{\varsigma}{\varsigma_H}\right)^2 = \left(\frac{k_H}{k}\right)^2,\\
\therefore ~~~ \frac{a_{\ast \ast}(k)}{a_\ast (k)} &=& \frac{a_\ast(k_H)}{a_\ast (k)} \times  \left[{\rm previous~interval's}~e^{r_k}\right].
\end{eqnarray}
Following in this step for this range and the next, we get
\begin{eqnarray}
r_k &=& \ln \left[ \left(\frac{k}{k_H}\right)^{\beta + 1} \left(\frac{k_H}{k_2}\right)^{\beta - 1} \left(\frac{k_2}{k_1}\right)^{\beta} \left(\frac{k_s}{k_1}\right)^{-\beta_s} \right], ~~k_E \leqslant k \leqslant k_H, \label{rk4}\\
r_k &=& \ln \left[ \left(\frac{k}{k_H}\right)^{\beta} \left(\frac{k_E}{k_H}\right) \left(\frac{k_H}{k_2}\right)^{\beta - 1} \left(\frac{k_2}{k_1}\right)^{\beta} \left(\frac{k_s}{k_1}\right)^{-\beta_s}\right], ~~ k \leqslant k_E. \label{rk5}
\end{eqnarray}
From Eqs.\eqref{rk1}-\eqref{rk5}, it can be seen that the reheating parameter $\beta_s$ affects the squeezing parameter for each range. Thus, $\beta_s$ affects the amplitude of the squeezed gravitational waves for the entire frequency ranges unlike the case without squeezing effect where $\beta_s$ actually affects only the reheating stage itself.\footnote{For details on the calculations of the amplitude without squeezing effect, one may refer to Ref.\cite{cqg}.}

Thus, both the squeezing parameters and squeezing angles vary for each frequency interval. The factor $\cos\phi_k (\varsigma)$ in Eqs.\eqref{hkbe} is an oscillatory function of time which reflects the oscillatory features in the power spectrum as a result of the squeezing effect. For long wavelength modes, $k < k_H$, $\cos \phi_k (\varsigma)$ is almost unity while it leads to oscillations for high frequency modes $k \gg k_H$.  Due to the squeezing effect, the stochastic primordial gravitational waves form a collection of standing waves with a non-stationary background.

\section{Amplitude and spectral energy density of primordial GWs}
In this section, we study  the amplitude and spectral energy density of the primordial gravitational waves in the squeezed vacuum state. The amplitude of  primordial gravitational waves can be determined with Eq.\eqref{hkbe}.

The fractional energy density of gravitational waves can be defined in terms of spectral energy density $\Omega_{gw} (\nu)$  as:
\begin{equation}\label{rhg}
\frac{\rho_{gw}}{\rho_c}=\int \Omega_{gw}(\nu) \frac{d\nu}{\nu},
\end{equation}
where $\rho_{gw}$ is the energy density of the gravitational waves and $\rho_c$ is the critical energy density of the universe. Since we are assuming spatially flat spacetime, the fractional energy density of primordial gravitational waves relative to the critical density of the universe must be less than 1.  The spectral energy density can be given in terms of the field as,
\begin{equation}
\Omega_{gw}(\nu)=\frac{\pi^2}{3} h^2(\nu) \left(\frac{\nu}{\nu_H}\right)^2.
\end{equation}
The wave number $k$ is proportional to the frequency $\nu$, so the ratios of the wave numbers can be replaced by the ratios of the frequencies. The Hubble frequency is $\nu_H= 1/l_H \simeq 2 \times 10^{-18}$ Hz. For other values of frequency $\nu$, we choose $\nu_E = 1.5 \times 10^{-18} $ Hz which is in the long wavelength regime, $\nu_2 = 117 \times 10^{-18}$ Hz and $\nu_s = 10^8$ Hz for definiteness, and $\nu_1 = 10^{10}$ Hz as it is the highest frequency at which the spectral energy density $\Omega_{gw} (\nu)$ in high frequency modes does not exceed the nucleosynthesis bound ($10^{-6}$). 

Eq.\eqref{qn} gives the initial normalized amplitude of the wave mode when the wave mode enters the long wavelength regime, at this instant its corresponding wavelength  $\lambda_i$ becomes equal to the Hubble radius. For this, the wavelength is,
\begin{equation}
\lambda_i = \frac{1}{b}l_0 \left(\frac{\nu_H}{\nu}\right)^{2+\beta}.
\end{equation}
Then,
\begin{equation}
h(k,\varsigma) = A \left(\frac{\nu}{\nu_H}\right)^{2+\beta},
\end{equation}
where $A = 8\sqrt{\pi} b\frac{l_{pl}}{l_0}$.
The wavelength of the gravitational wave mode at this instant must be greater than the Planck length $l_{pl}$. Thus,
\begin{equation}
b\frac{l_{pl}}{l_0}\left(\frac{\nu}{\nu_H}\right)^{2+\beta} < 1.
\end{equation}
Observational constraints give a restriction \cite{lp2},
\begin{equation}
b\frac{l_{pl}}{l_0} \approx 10^{-6}.
\end{equation}
Thus, at the highest frequency $\nu = \nu_1$, we get
\begin{equation}
\left(\frac{\nu_1}{\nu_H}\right)^{2+\beta} < 10^6,
\end{equation}
which provides the upper bound on $\beta$ which is $-1.77$, i.e., $\beta < -1.77$. Thus, using the above equations with Eq.\eqref{lo}, and $l_H/l_{pl}=9.276 \times 10^{59}$ (after converting them into same units), the allowed values of $\beta_s$ with respect to $\beta =  -1.8, -1.9, -2.0$ are calculated as $\beta_s =  0.598, -0.538, -1.676$ respectively. Note that this explicit calculation of $\beta_s$ defies the constraint $1+\beta_s > 0$ when it comes to squeezing as $\beta_s$ now affects the entire spectrum \cite{lp2}. The new constraint for $\beta_s$ is $\beta_s <1$ or $|1+\beta_s|>0$.

Using Eq.\eqref{rhg}, the consistency of these models in the absence ($r_k = 0$) and presence ($r_k \neq 0$) of the squeezing effect can be examined by taking integration over all frequencies.\footnote{Note that for $r_k = 0$, $h(\nu)$ is calculated not by following Eq.\eqref{hkbe} but by following Ref.\cite{cqg}.}

\noindent
For $\beta = -1.8$,
\begin{eqnarray}
\frac{\rho_{gw}}{\rho_c} &=& 9.29 \times 10^{-4}, ~~r_k = 0, \nonumber \\
\frac{\rho_{gw}}{\rho_c} &=& 6.83 \times 10^{-3}, ~~r_k \neq 0.
\end{eqnarray}

\noindent
For $\beta = -1.9$,
\begin{eqnarray}
\frac{\rho_{gw}}{\rho_c} &=& 1.989 \times 10^{-6}, ~~r_k = 0, \nonumber \\
\frac{\rho_{gw}}{\rho_c} &=& 1.78 \times 10^{-5}, ~~~r_k \neq 0.
\end{eqnarray}

\noindent
For $\beta = -2.0$,
\begin{eqnarray}
\frac{\rho_{gw}}{\rho_c} &=& 7.59 \times 10^{-7}, ~~r_k = 0, \nonumber \\
\frac{\rho_{gw}}{\rho_c} &=& 9.97 \times 10^{-6}, ~~r_k \neq 0.
\end{eqnarray}
Hence the models discussed are not ruled out both in the absence and presence of squeezing effect. If detected, the observed features would hopefully reveal whether or not the primordial gravitational waves are in the squeezed vacuum state.

\begin{figure}
\centering
\subfloat[]
{\includegraphics[scale=0.42]{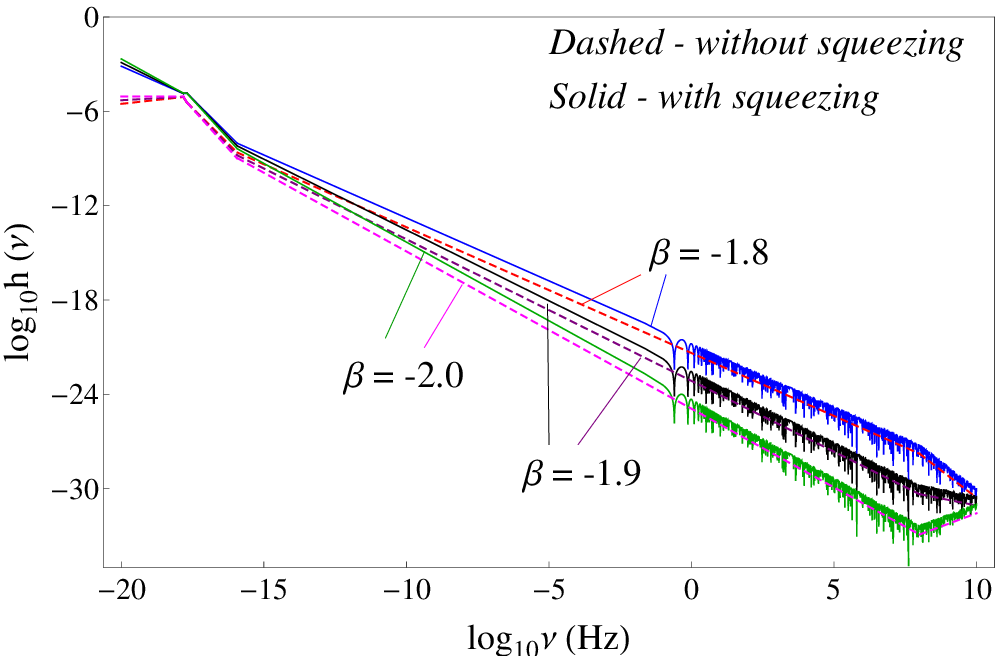}
\label{amp_all1}}
\subfloat[]
{\includegraphics[scale=0.37]{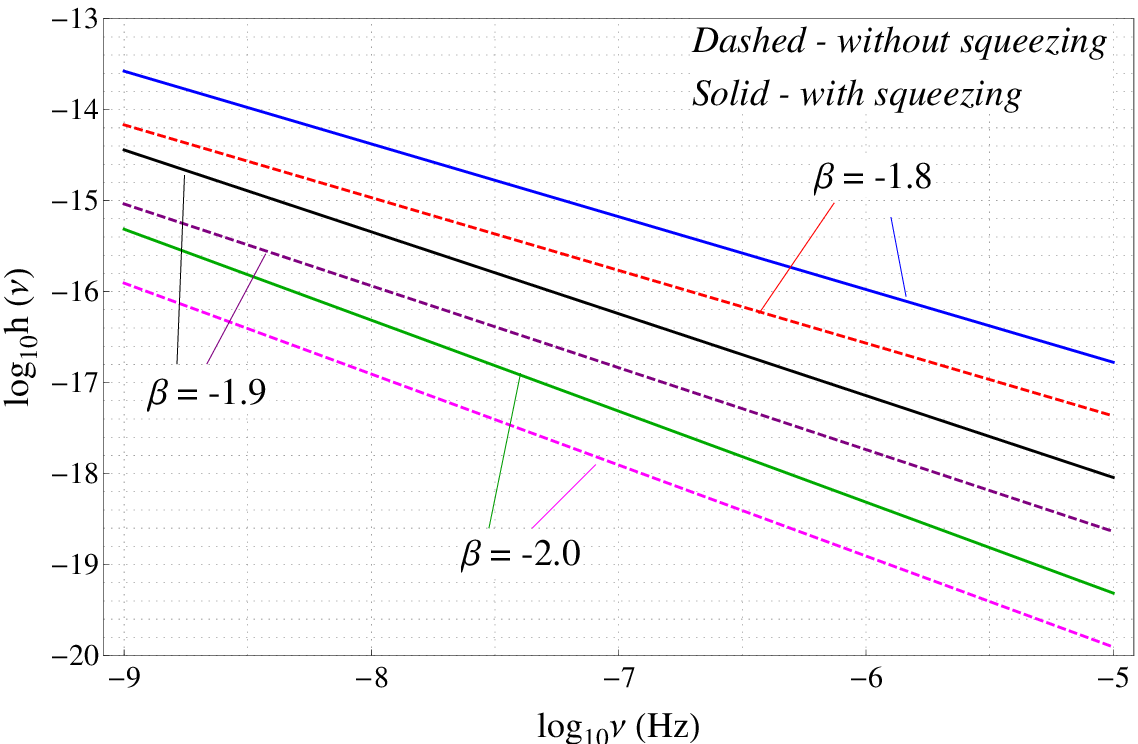}
\label{amp_s}}
\caption{Amplitude of stochastic background of primordial gravitational waves as a function of frequency for $\beta = -1.8$, $\beta = -1.9$, $\beta = -2.0$ in the presence  and absence of squeezing effect. \ref{amp_s} is just a zoom in into parts of the amplitude in \ref{amp_all1}.}
\label{f1}
\end{figure}

\begin{figure}
\centering
\subfloat[]
{\includegraphics[scale=0.415]{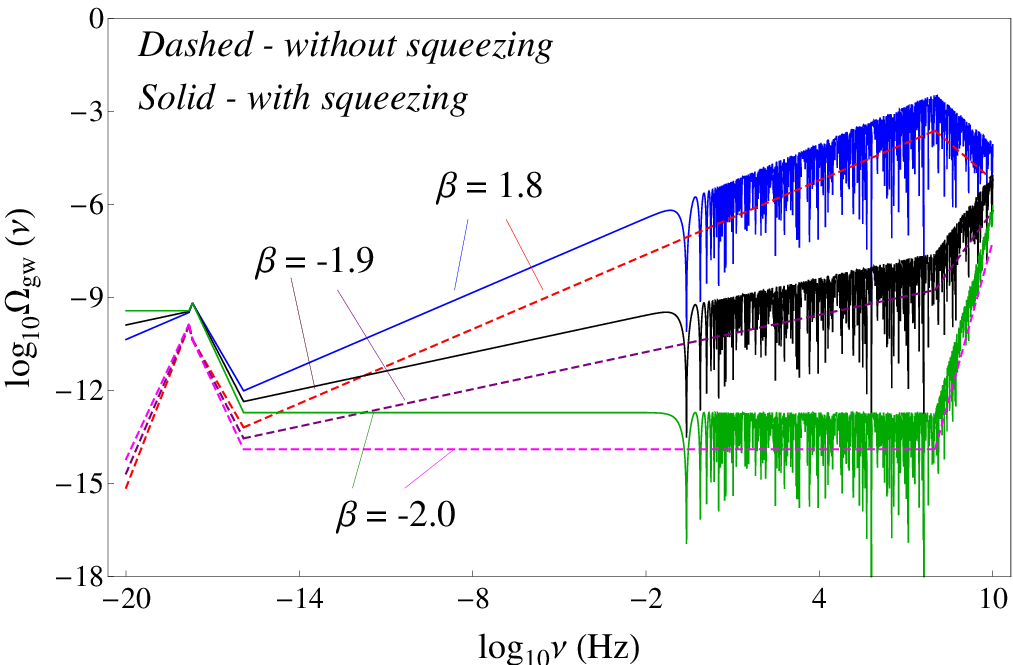}
\label{omega_all}}
\subfloat[]
{\includegraphics[scale=0.37]{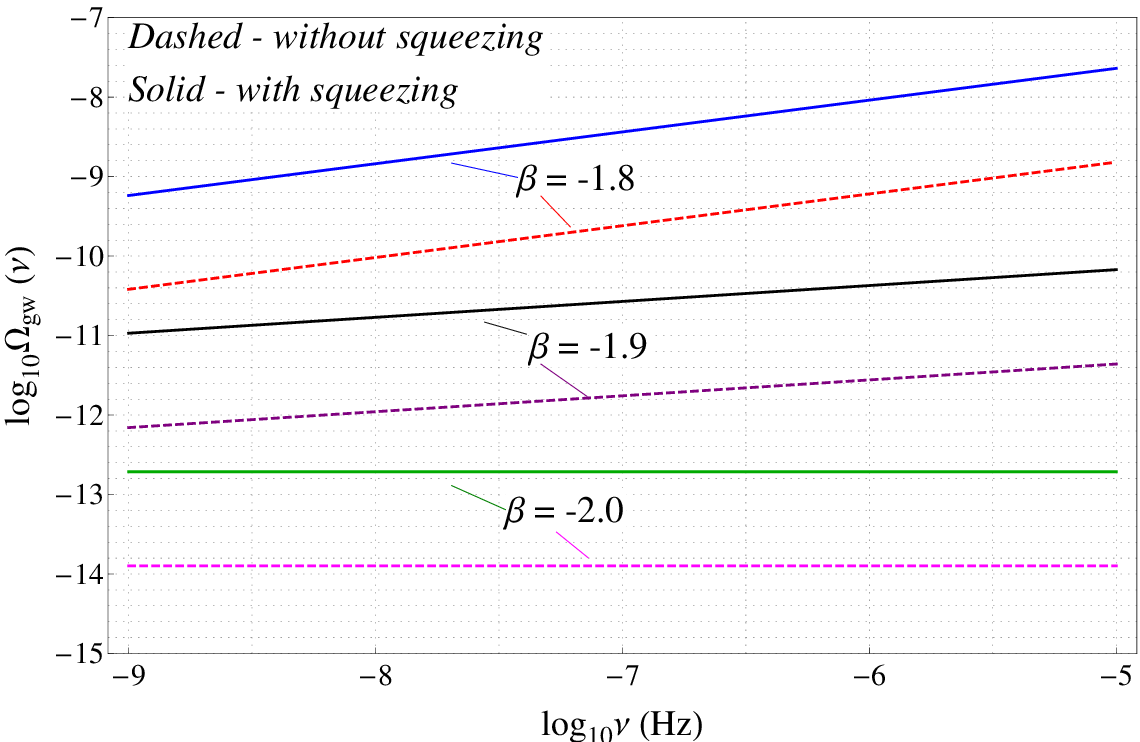}
\label{omega_s}}
\caption{Spectral energy density of stochastic background of primordial gravitational waves as a function of frequency for $\beta = -1.8$, $\beta = -1.9$, $\beta = -2.0$ in the presence and absence of squeezing effect. \ref{omega_s} is a zoom in into parts of the spectral energy density in \ref{omega_all}.}
\label{f2}
\end{figure}

\begin{figure}
\begin{center}
\includegraphics[scale=0.3]{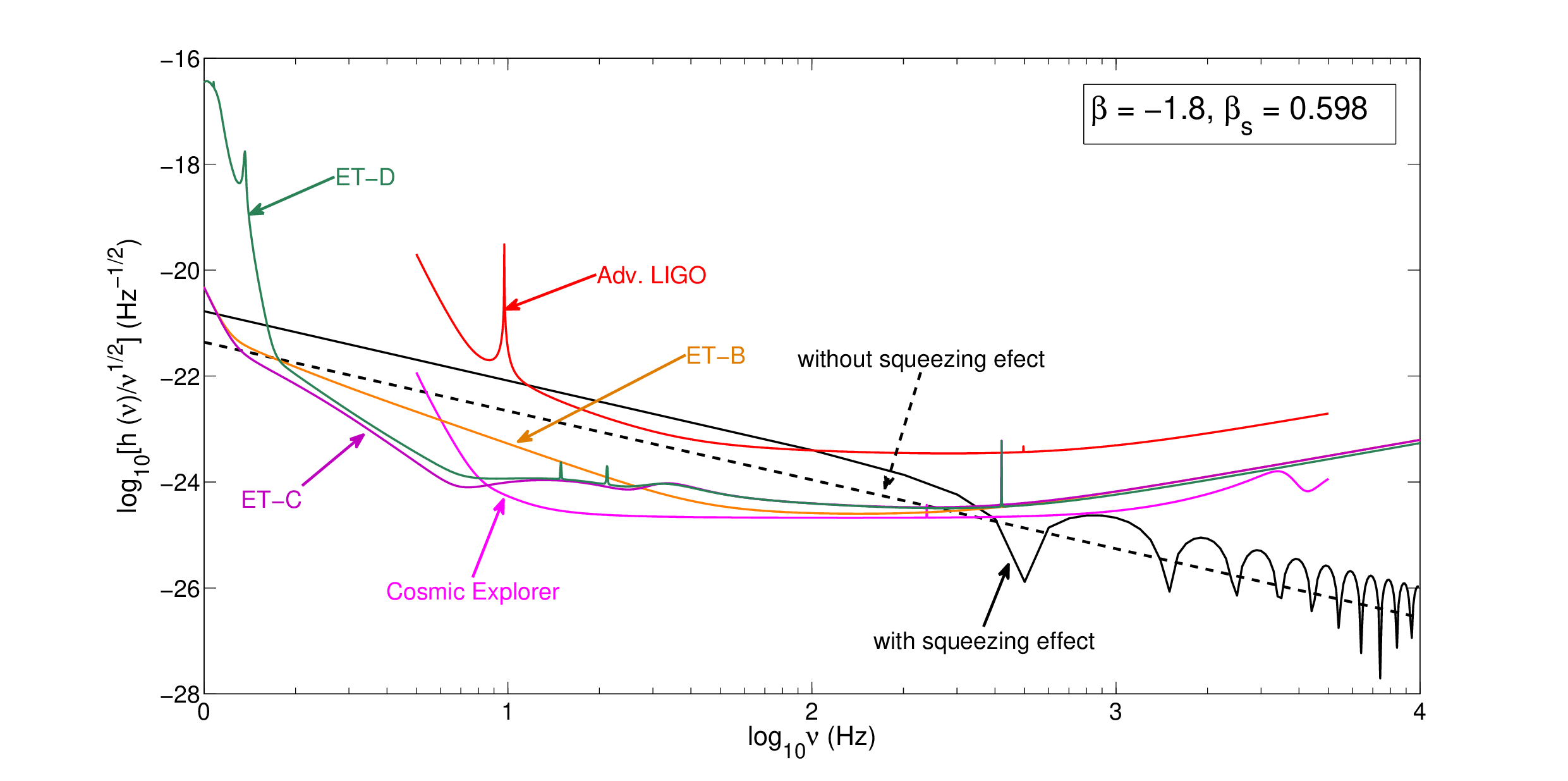}
\caption{Amplitudes of gravitational waves as a function of frequency for $\beta = -1.8$ compared with sensitivity curves of Advanced LIGO, Cosmic Explorer and Einstein Telescope.}\label{f3}
\end{center}
\end{figure}

\begin{figure}
\begin{center}
\includegraphics[scale=0.3]{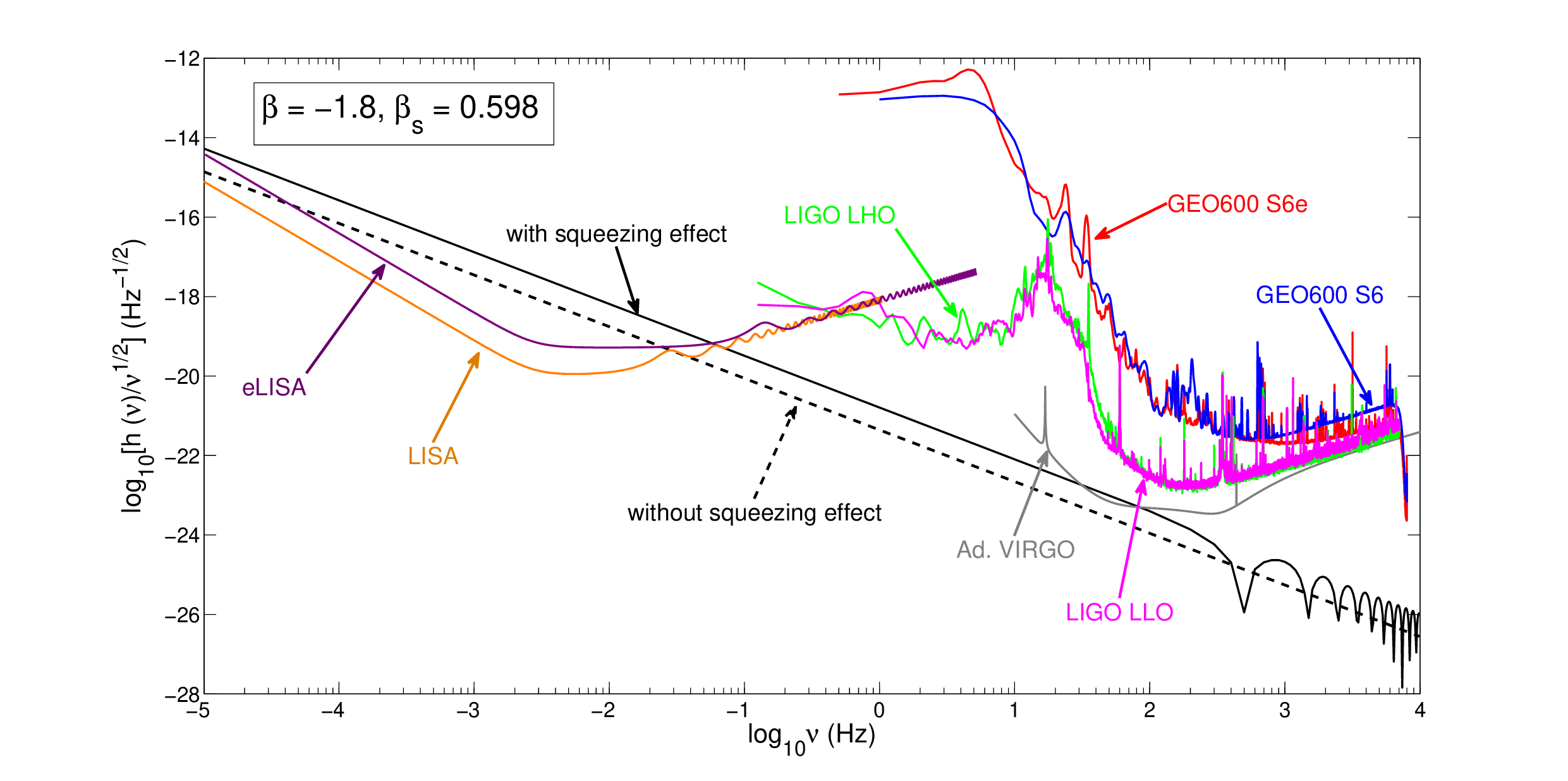}
\caption{Amplitudes of gravitational waves as a function of frequency for $\beta = -1.8$ compared with sensitivity curves of Advanced VIRGO, GEO-600, LIGO S6, LISA and eLISA.}\label{f4}
\end{center}
\end{figure}

\begin{figure}
\begin{center}
\includegraphics[scale=0.3]{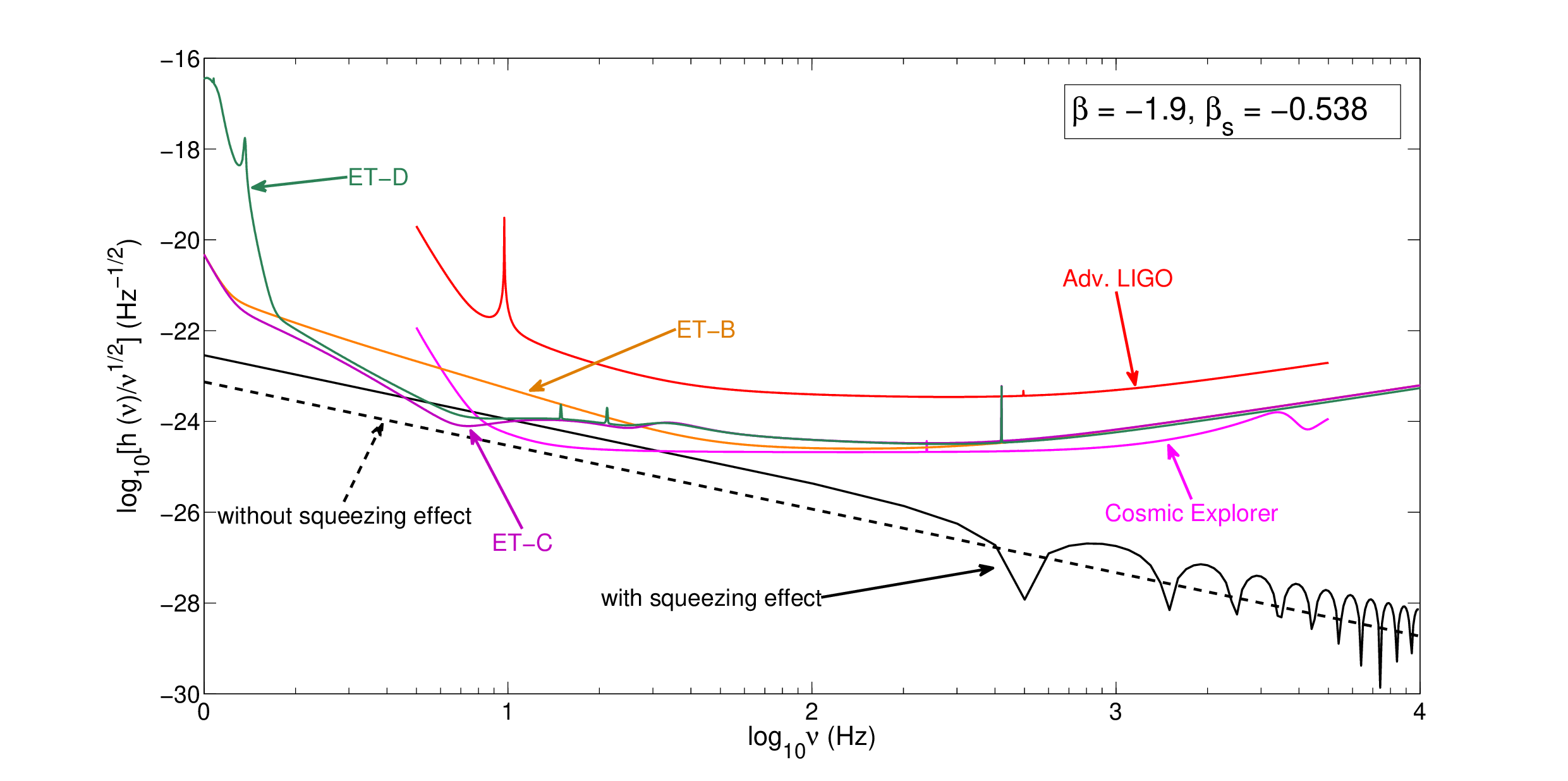}
\caption{Amplitudes of gravitational waves as a function of frequency for $\beta = -1.9$ compared with sensitivity curves of Advanced LIGO, Cosmic Explorer and Einstein Telescope.}\label{f5}
\end{center}
\end{figure}

\begin{figure}
\begin{center}
\includegraphics[scale=0.3]{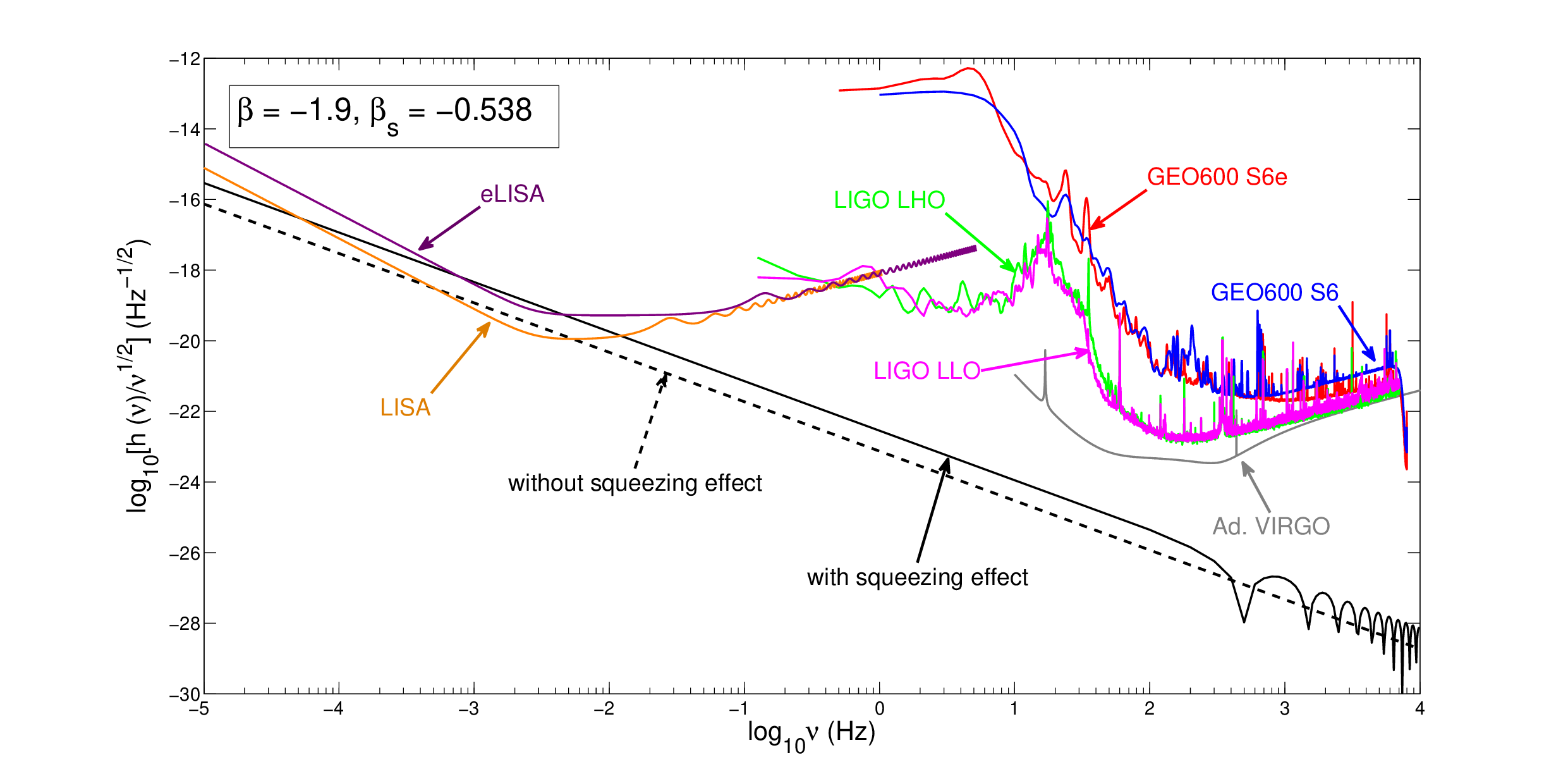}
\caption{Amplitudes of gravitational waves as a function of frequency for $\beta = -1.9$ compared with sensitivity curves of Advanced VIRGO, GEO-600, LIGO S6, LISA and eLISA.}\label{f6}
\end{center}
\end{figure}

\begin{figure}
\begin{center}
\includegraphics[scale=0.3]{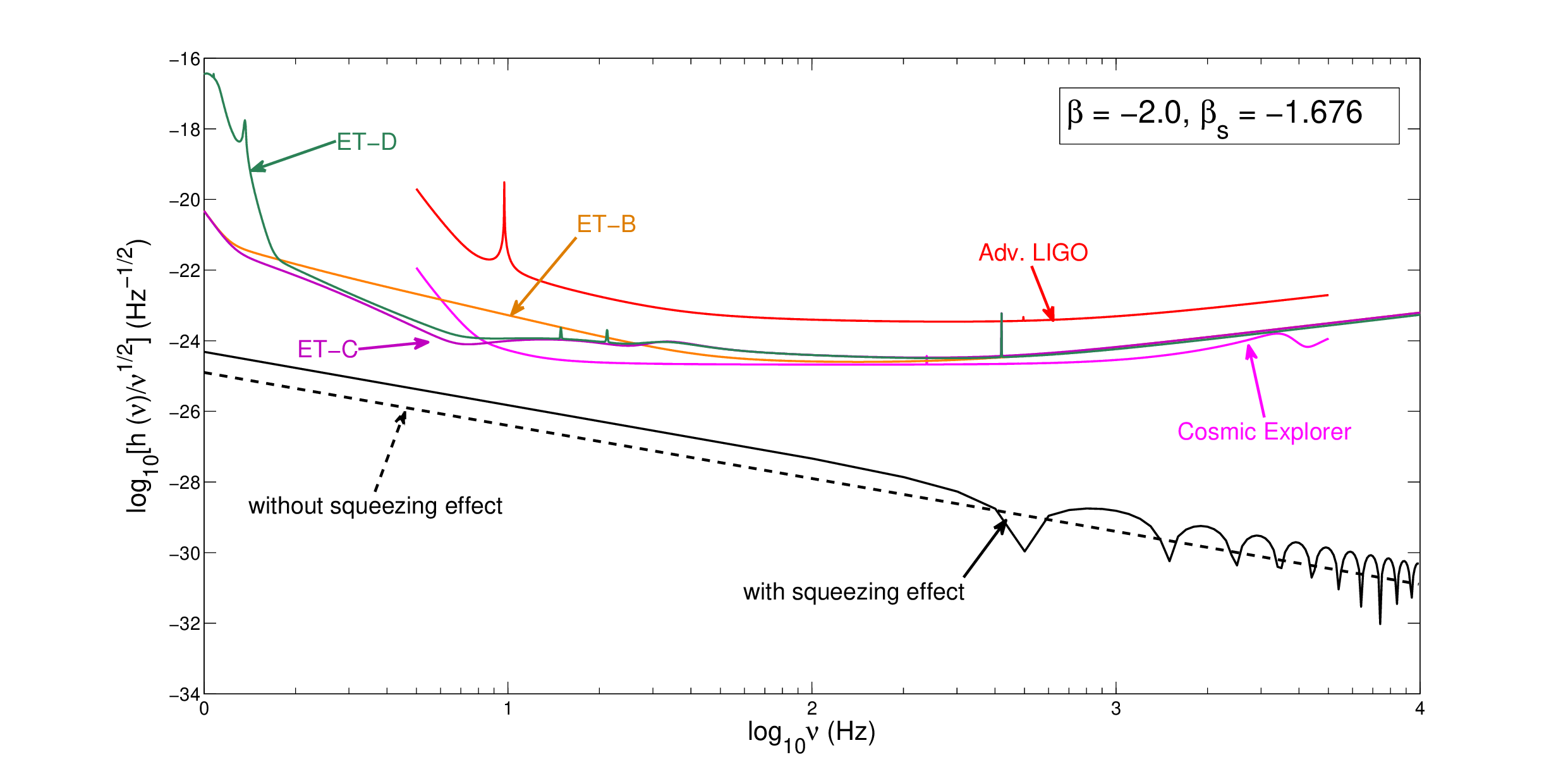}
\caption{Amplitudes of gravitational waves as a function of frequency for $\beta = -2.0$ compared with sensitivity curves of Advanced LIGO, Cosmic Explorer and Einstein Telescope.}\label{f7}
\end{center}
\end{figure}

\begin{figure}
\begin{center}
\includegraphics[scale=0.3]{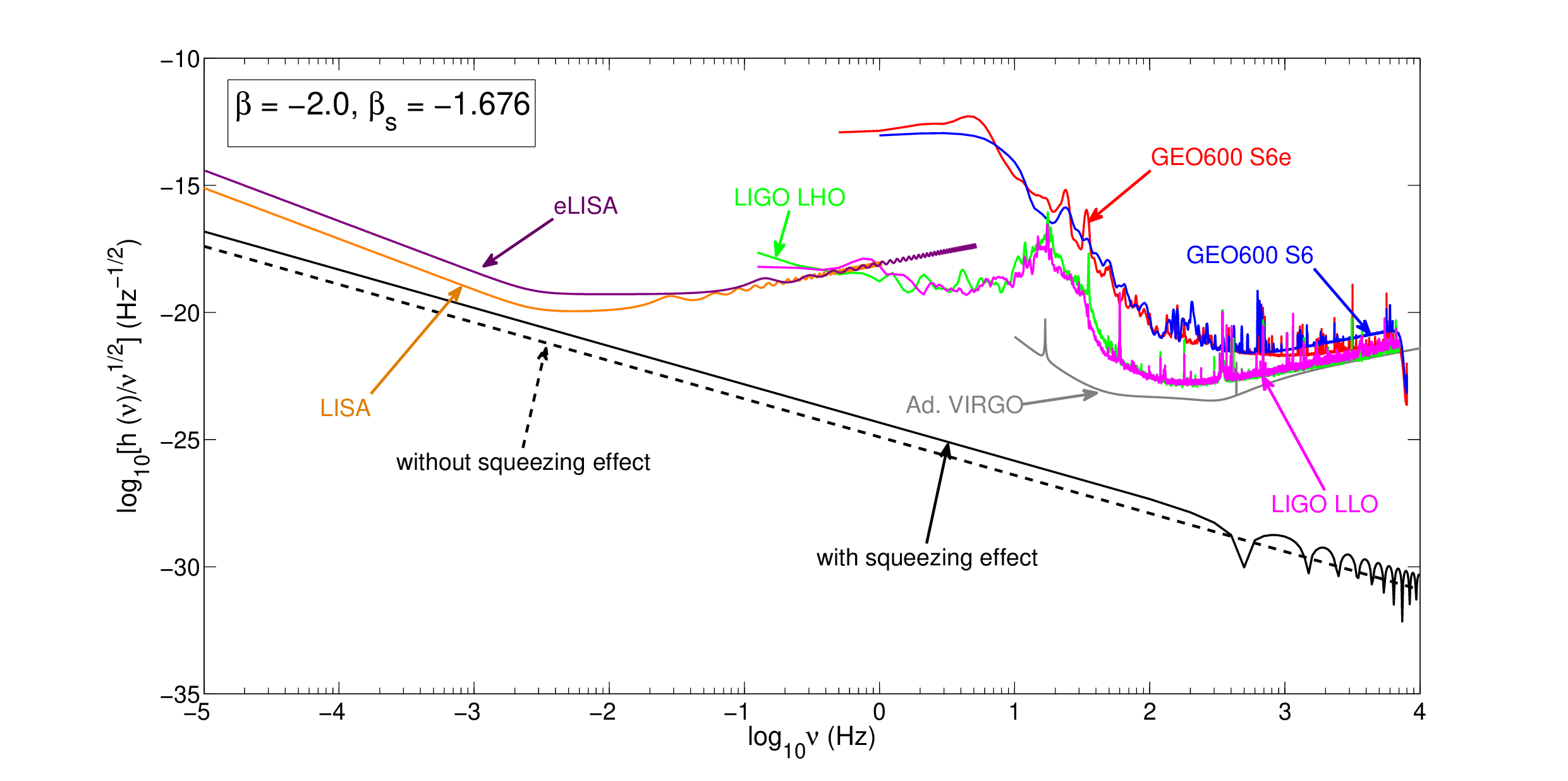}
\caption{Amplitudes of gravitational waves as a function of frequency for $\beta = -2.0$ compared with sensitivity curves of Advanced VIRGO, GEO-600, LIGO S6, LISA and eLISA.}\label{f8}
\end{center}
\end{figure}

The amplitude and spectral energy density for primordial gravitational waves are presented  in Figs.\ref{f1} and \ref{f2} respectively. The amplitude $h(\nu)$ in the presence of squeezing effect is obtained with the help of Eqs.\eqref{hkbe}, \eqref{sqa1}-\eqref{sqa5}, \eqref{rk1}-\eqref{rk3} and \eqref{rk4}-\eqref{rk5} after substituting the wave numbers by frequencies. The analysis is done for various models of expanding universe, that is, $\beta = -1.8$, $\beta = -1.9$ and $\beta = -2.0$. Color assignments are same for these figures. For $\beta = -1.8$, we used blue and red for spectrum with and without squeezing respectively. For $\beta = -1.9$, we used black and purple for spectrum with and without squeezing respectively. Finally, for $\beta = -2.0$, we used green and magenta for spectrum with and without squeezing respectively. For these figures, we used dashed lines to denote the spectrum without squeezing effect for clarity. From Figs.\ref{f1}, it can be checked with the background grid lines that for each model, the enhancement of the amplitude due to squeezing effect is $\sim 60 \%$ at $\nu > 10^{-16}$ Hz compared to the amplitude without squeezing effect. Similarly, it can be seen from Figs.\ref{f2} that the enhancement of the spectral energy density due to squeezing effect is $\sim 120 \%$ at $\nu > 10^{-16}$ Hz. 

For each of these models, the obtained amplitude of gravitational waves (Figs.\ref{f3}-\ref{f8}) can be used to  examine the possibility of their detection with sensitivity curves from Advanced LIGO (red), Cosmic Explorer (magenta), Einstein Telescope (B - orange, C - purple, D - dark green) in one set of figures, and Advanced VIRGO (gray), GEO-600 (S6e - red, S6 - blue), LIGO S6 (LHO - green and LLO - magenta), LISA (orange) and eLISA (dark purple) in another set. In Figs.\ref{f3}-\ref{f8}, we used solid black lines for squeezed spectrum and dashed black for non-squeezed spectrum. These comparisons are made with respect to root mean square amplitude per root Hertz plot ($h(\nu)/\sqrt{\nu}$), also known as the amplitude spectral density.

\section{Results and discussion}
We have studied the stochastic background of primordial gravitational waves for various expanding universe models which show slight deviations from the exact de Sitter expansion, $\beta = -1.8$ and $\beta = -1.9$, and for the exact de Sitter case itself, $\beta = -2.0$. We have presented the calculated squeezing parameters and angles for each frequency range. We found that the phenomenon of squeezing affects the entire spectrum. We also found that through the squeezing effect, the reheating parameter $\beta_s$ affects the entire spectrum, unlike the case in the absence of squeezing where $\beta_s$ affects only the reheating stage.

We observe strong oscillating features towards higher frequency due to squeezing effect. As the wave enters the long wavelength regime (lower frequency), the oscillation decreases. The oscillations arise due to the interaction of the gravitational wave field (which acts as oscillator) with the strong and variable gravitational field of the early universe which amplifies the zero-point quantum fluctuations of gravitational waves associated with the vacuum state energy $\omega/2$ that ended up in the stochastic background.

There is enhancement in the amplitude for each model due to the squeezing effect as compared to models in the absence of squeezing effect. A pair of one-mode squeezed waves traveling in opposite direction interfere to form a standing wave of enhanced amplitude. The strength and variability of the gravitational field of the very early universe and the very nature of the squeezed quantum states determine the r.m.s value of the field and its statistical properties. The squeezing parameter grows with time. The squeezed variances of phase and the associated standing wave pattern of the field make the spectrum an oscillating function of wave number or frequency for a fixed moment of time while the large variances of amplitude lead to enhancement in the spectrum. The energy of the gravitational wave field has been increased at the expense of the variable gravitational field.

The enhancement in the amplitude is $\sim 60\%$ for frequency $>10^{-16}$ Hz while the enhancement in the energy is $\sim 120\%$ for frequency $> 10^{-16}$ Hz for each model. However, while the amplitude  for the models without squeezing effect remains almost constant when the wave enters the long wavelength regime, the amplitude for the models in the presence of squeezing effect keeps on increasing. This is due to the fact that the variance of the gravitational wave mode's phase is strongly squeezed while its amplitude is strongly increased, a feature which continues upto the very end of the amplifying regime \cite{lp1, lp2}. This is in accordance with the uncertainty principle. The squeezing parameter remains constant for a brief period due to adiabatic conditions around $ (10^{-16} \sim 10^{-18})$ Hz and again increases with the accelerating expansion.
 
We also examine the possibility of detection of the primordial gravitational waves for each model for various detectors. Figs.\ref{f3} and \ref{f4} show that for the model $\beta = -1.8$ both in the absence and presence of squeezing effect, the field lies within the sensitivity range of Einstein Telescope, Cosmic Explorer and LISA (original design and evolved design). Additionally, the amplitude in the presence of squeezing effect also lies within the sensitivity range of advanced LIGO and advanced VIRGO. On the other hand, according to Figs.\ref{f5}-\ref{f6}, the field for the model $\beta = -1.9$ in the presence of squeezing effect lies within the sensitivity range of Einstein Telescope, Cosmic Explorer, eLISA and LISA. The gravitational wave field for $\beta = -2.0$ both in the absence and presence of squeezing effect (Figs.\ref{f7}-\ref{f8}) lies below the sensitivity curves of all the mentioned detectors. Thus, the prospect of testing the case for $\beta = -2.0$ is currently out of scope.

\section{Conclusions}
Primordial gravitational waves generated in the very early universe  have traversed the universe since their generation and their wavelengths have stretched significantly during their course of travel with the expansion of the universe, the time-dependent expansion of the wavelength being $\lambda (\varsigma) = 2\pi a(\varsigma)/k$.  By calculating the values of $l_0 = 6.0968 \times 10^{-18}$ m and $|\varsigma_1| = 6.07787 \times 10^{-23}$, and using them to find the corresponding scale factor, it can be found that the wavelength at the end of inflation is $\sim 7.6 \times 10^{-10}$ m, calculations being made with $\beta = -1.8$. Calculations for $\beta = -2.0$ yield the wavelength $\sim 2.22 \times 10^{-11}$ m. Then, with straightforward calculations, i.e., $\lambda \propto \nu$ since gravitational waves travel with the speed of light, the wavelength corresponding to present-day Hubble radius can be found as $\sim 1.5 \times 10^{26}$ m.

The energy and strength of primordial gravitational waves generated during inflation have obviously diminished significantly with the increase in wavelength. Therefore they are believed to be the most difficult form of gravitational waves to detect. Hence, despite the fact that some of the amplitudes in the present analysis lie within the sensitivity range of some detectors, this type of gravitational waves is believed to be very quiet and extremely difficult to detect. These waves are believed to form a stochastic background with standing wave pattern whose spectrum depends on the evolutionary stages of the universe. Also, since these waves are believed to be generated quantum mechanically due to quantum fluctuations in the very early universe, they are expected to be in a specific quantum state called the squeezed vacuum state. Due to this effect, the amplitudes of these waves are modified throughout the amplifying regime, see  Figs.\ref{f1}-\ref{f8}.  If detected, these waves are expected to produce a continuous static noise. The non-stationary property of these primordial gravitational waves could provide a unique and distinguishing signature from a background of stationary stochastic background created from other sources or processes.

We would like to comment further on the prospect of detectability of squeezed states of primordial gravitational waves. The impossibility to distinguish between observable signals imparted due to squeezing and non-squeezing was discussed some time back where it was further mentioned that the non-stationary feature is statistical which needs measurement as long as the age of the universe itself \cite{det}. However, there have recently been studies on the potential detectability via quantum noise. It has been shown that quantum mechanical treatment of gravitational field can induce fluctuations or noise in the lengths of the arms of gravitational wave detectors where the characteristics of the noise depends on the quantum state of the gravitational field \cite{qn1,qn2}. Such noise is very small for coherent states, but can be greatly enhanced especially in squeezed states and are potentially detectable. An experimental setup for this case has further been theoretically proposed \cite{qn2}. Another study on the detectability of quantum noise in the squeezed state from the induced gravitons has shown that while gravitational waves from all sources produce quantum noise if they arrive at the detectors, only the gravitational wave background from the early universe in the squeezed states could produce enough energy density which would be able to be detected while the quantum noise in the non-squeezed state is much smaller \cite{qn4}. This is also supported by the present evaluation where the energy density is larger by $\sim 120\%$ as compared to the energy density without squeezing.

\appendix
\section{Hamiltonian and equation of motion for one-mode squeezed state}
The Hamiltonian associated with the Heisenberg equations of motion in Eqs.\eqref{Hsa} and \eqref{Hsb} is of the form,
\begin{equation}
H_{gw} = \frac{1}{2}\left[p^2 + \frac{a'}{a}(\mu_k p + p \mu_k) + k^2 \mu_k^2 \right],
\end{equation}
where $p$ is the canonically conjugated momentum, $p=\mu_k'-(a'/a)\mu_k$. In the quantum treatment, $\mu_k$ and $p$ are operators satisfying the commutation relation $[\mu_k, p] = i$. The associated annihilation and creation operators are respectively,
\begin{eqnarray}
c_k = \sqrt{\frac{k}{2}}\left(\mu_k + i \frac{p}{k}\right), ~~~~~~~~~~ c_k^\dagger = \sqrt{\frac{k}{2}}\left(\mu_k - i \frac{p}{k}\right),
\end{eqnarray}
which satisfy $[c_k,c_k^\dagger] =1$. Then, the Hamiltonian acquires the form,
\begin{equation}
H_{gw} = kc_k^\dagger c_k + \sigma(\varsigma)c_k^{\dagger 2} + \sigma^\ast (\varsigma) c_k^2,
\end{equation}
where the coupling function is $\sigma (\varsigma) = ia'/2a$. Using this Hamiltonian in Eqs.\eqref{Hsa} and \eqref{Hsb}, the Heisenberg equations of motion have the form,
\begin{eqnarray}
i\frac{dc_k}{d\varsigma} &=& kc_k + i \frac{a'}{a}c_k^\dagger, \label{h1} \\
-i\frac{dc_k^\dagger}{d\varsigma}  &=& kc_k^\dagger - i \frac{a'}{a}c_k. \label{h2}
\end{eqnarray}
These equations have the solutions,
\begin{eqnarray}
c_k (\varsigma) &=& u_k (\varsigma)c_k(0) + v_k (\varsigma) c_k^\dagger (0), \label{cs1} \\
c_k^\dagger (\varsigma) &=& u_k^\ast (\varsigma) c_k^\dagger (0) + v_k^\ast (\varsigma) c_k (0), \label{cs2}
\end{eqnarray}
where $c_k(0)$ and $c_k^\dagger(0)$ are the initial values of $c_k(\varsigma)$ and $c_k^\dagger(\varsigma)$. Differentiating Eqs.\eqref{cs1} and \eqref{cs2} w.r.t $\varsigma$ yields,
\begin{eqnarray}
\frac{dc_k}{d\varsigma} &=&  \frac{d u_k}{d\varsigma} c_k(0) + \frac{dv_k}{d\varsigma} c_k^\dagger (0), \label{csa} \\
\frac{dc_k^\dagger}{d\varsigma} &=&  \frac{d u_k^\ast}{d\varsigma} c_k^\dagger(0) + \frac{dv_k^\ast}{d\varsigma} c_k (0). \label{csb}
\end{eqnarray}
Equating Eqs.\eqref{csa} and \eqref{csb} with \eqref{h1} and \eqref{h2} and collecting the coefficients of the annihilation and creation operators, we get
\begin{eqnarray}
i\frac{du_k}{d\varsigma} &=& ku_k (\varsigma) + i\frac{a'}{a} v^{\ast}_k (\varsigma), \nonumber \\
 i\frac{dv_k}{d\varsigma} &=& kv_k (\varsigma) + i\frac{a'}{a} u^{\ast}_k (\varsigma),
\end{eqnarray}
where $|u_k|^2-|v_k|^2 =1$, $u_k(0)=1$, $v_k(0)=0$. This description based on $c_k$ and $c_k^\dagger$ operators corresponds to one-mode squeezed states and standing waves.

\section*{Acknowledgment}
N M and  P K S would like to thank  DST-SERB, New Delhi for financial support. The authors also thank the reviewers for their comments and suggestions.

\end{document}